\documentclass{mn2e}
\usepackage{epsf}
\usepackage{times}
\newcommand{\etal}{{et al}\/.}
\begin{document}
\title[Radio galaxies at 90 GHz]{The properties of powerful radio
sources at 90 GHz}
\author[M.J.~Hardcastle \& L.W. Looney]{M.J.\ Hardcastle$^{1,2}$ and
L.W.\ Looney$^3$ \\
$^1$ School of Physics,
  Astronomy and Mathematics, University of
Hertfordshire, College Lane, Hatfield, Hertfordshire AL10 9AB\\
$^2$ Department of Physics, University of Bristol, Tyndall Avenue,
Bristol BS8 1TL\\
$^3$ Department of Astronomy, University of Illinois, 1002 West Green
Street, Urbana, IL 61801, U.S.A.}
\maketitle
\begin{abstract}
We have observed a small sample of powerful double radio sources
(radio galaxies and quasars) at frequencies around 90 GHz with the
BIMA millimetre array, with the intention of constraining the resolved
high-frequency spectra of radio galaxies. When combined with other
sources we have previously observed and with data from the BIMA
archive, these observations allow us for the first time to make
general statements about the high-frequency behaviour of compact
components of radio galaxies -- cores, jets and hotspots. We find that
cores in our sample remain flat-spectrum up to 90 GHz; jets in some of
our targets are detected at 90 GHz for the first time in our new
observations; and hotspots are found to be almost universal, but show
a wide range of spectral properties. Emission from the extended lobes
of radio galaxies is detected in a few cases and shows rough
consistency with the expectations from standard spectral ageing
models, though our ability to probe this in detail is limited by the
sensitivity of BIMA. We briefly discuss the prospects for radio-galaxy
astrophysics with ALMA.
\end{abstract}

\section{Introduction}
\label{intro}

Although they are the dominant radio source population in the sky at
low radio frequencies, steep-spectrum extragalactic radio sources
(radio galaxies and lobe-dominated radio-loud quasars) have been
relatively poorly studied at millimetre wavelengths (corresponding to
frequencies $\ga 100$ GHz), although a good deal of work has been done
on integrated flux density measurements of their flat-spectrum,
jet-dominated counterparts (e.g.\ Edelson 1987). Where information on
steep-spectrum radio sources at mm wavelengths is available in the
literature (e.g.\ Steppe \etal\ 1988, 1995) it tends to be restricted
to tables of integrated flux densities. This is presumably partly due
to the intrinsic difficulty of mm-wave observations compared to those
at 1--10 GHz, partly because the steep spectrum of most components of
radio galaxies makes them harder to detect at higher frequencies, and
partly because of a perception that there is little additional
information to be gained at high frequencies.

High-frequency imaging observations of radio galaxies and quasars can
have some astrophysical importance, however. In early work on the
spectrum of the large-scale radio emission from powerful radio
galaxies (FRIIs, Fanaroff \& Riley 1974), it was found that their
spectra, measured at centimetre wavelengths, systematically steepen
with distance from the hotspots, inferred to be the sites of particle
acceleration at each end of the source (e.g. Myers \& Spangler 1985,
Alexander \& Leahy 1987). The steepening was interpreted as evidence
for `synchrotron ageing', whereby the synchrotron energy loss process
depletes high-energy electrons in the source (e.g. Pacholczyk 1970,
Jaffe \& Perola 1973, Leahy 1991). If observations of spectral
steepening in FRIIs are really a result of synchrotron ageing, then in
principle resolved images of the radio spectra allow us to measure
plasma age as a function of position in the source, which in turn
allows the determination of quantities like total source age and
expansion speed. However, the interpretation of the spectral
steepening in FRIIs in terms of synchrotron ageing has been questioned
more recently. Most dramatically, Katz-Stone, Rudnick \& Anderson
(1993) showed that the detailed centimetre-wave spectrum of the
extended `lobe' emission in the powerful FRII source Cygnus A
(3C\,405) is not consistent with any conventional ageing model,
arguing instead that the observations are consistent with a single,
curved electron energy spectrum seen in a varying magnetic field; this
would imply that ageing is in fact a negligible effect in the lobes.
Introducing inhomogeneities into the field strength or electron
population also seems likely to undermine `traditional' spectral
ageing. Rudnick (1999) gives a summary of the theoretical and
observational problems faced by ageing models.

One important constraint on all models would be observations at
wavelengths shorter than the centimetre wavelengths typically used in
observations of radio galaxies. The effects of ageing become
greater as we go to shorter wavelengths; in particular, there should
be a very strong difference between the short-wavelength spectra of
regions near and far from the hotspots in the standard picture.
However, only a few powerful radio galaxies have been imaged at these short
wavelengths. Cygnus A, the brightest such object in the sky, has been
observed with the Berkeley Illinois Maryland Association millimetre-wave
interferometer (BIMA) on several occasions (Wright \& Birkinshaw 1984, Wright \&
Sault 1993) and with the IRAM 30-m telescope at 230 GHz (Salter \etal\
1989). Wright \& Sault (1993) commented briefly on
the global spectrum of the lobes of Cygnus A at $\sim 90$ GHz, but
their result does not tell us anything about the {\it variation} of the
centimetre-to-millimetre spectrum of the source as a function of
position. More recently, we used BIMA to observe the radio
galaxies 3C\,123 and 3C\,20 (Looney \& Hardcastle 2000, Hardcastle \&
Looney 2001) at similar frequencies. We noted that 
the integrated spectrum of a part of the northern lobe of 3C\,123 was
inconsistent with an ageing model, leading us to suggest that
particle acceleration might be ongoing in that region. An upper limit
on the flux density of the southern lobe was consistent with ageing
models, however. In 3C\,20, the centimetre-to-millimetre spectrum of a
part of the eastern lobe was very consistent with spectral ageing models,
in contrast to the results of Katz-Stone \etal\ 

In this paper we combine the results of this and other earlier work
with new observations of a small sample of FRII radio sources using
BIMA, taken in configurations intended to allow us to
separate lobes from core, jet and hotspot emission. When combined with
observations with the NRAO Very Large Array (VLA) and the
Multi-Element Radio Linked Interferometer Network (MERLIN) the resulting
database of objects allows us to make some general statements about
the high-frequency spectral properties of components of FRIIs and
about the viability of standard spectral ageing models. We also
comment on the prospects for observations of radio galaxies with the
Atacama Large Millimeter Array (ALMA).

J2000.0 co-ordinates are used throughout, and spectral index $\alpha$
is defined in the sense $S \propto \nu^{-\alpha}$. Resolutions quoted
are the major $\times$ minor FWHM of elliptical restoring Gaussians
(the Gaussian is circular if only one number is quoted) and the
position angles quoted are the angle between the major axis and north,
defined in the sense that positive angles indicate a rotation eastwards.

\section{The observations}

Our aim in making new observations was to detect the lobe-related
emission seen in our observations of 3C\,123 and 3C\,20. Accordingly,
we selected for observation FRII radio sources with high surface
brightness, so as to maximize the prospects of detecting the
comparatively low-surface-brightness lobes with BIMA. We chose our
targets from the 3CRR sample (Laing, Riley \& Longair 1983). After
excluding compact-steep-spectrum sources, which BIMA cannot image
usefully given their small angular size, and core-dominated objects,
we obtained the sample listed in Table \ref{sample}. The two sources
we had previously observed, 3C\,20 and 3C\,123, would also have been
selected from 3CRR by these criteria. 3C\,405, our final object, is
excluded from 3CRR because of its low Galactic latitude, but is of
course of very high surface brightness at low frequency. Data for
3C\,405, the B- and C-array components of the mosaiced observations
discussed by Wright, Chernin \& Forster (1997), were kindly supplied
by Mel Wright.

\begin{table}
\caption{The surface-brightness-selected sample of FRII radio sources}
\label{sample}
\begin{tabular}{llrrr}
\hline
Source&Newly&$z$&LAS&178-MHz flux\\
&observed?&&(arcsec)&density (Jy)\\
\hline
3C\,20&N&0.174&53.6&46.8\\
3C\,123&N&0.218&38&206.0\\
3C\,220.1&Y&0.61&35&17.22\\
3C\,295&Y&0.461&6&91.0\\
3C\,388&Y&0.091&50&26.81\\
3C\,401&Y&0.201&23.6&22.78\\
3C\,405&N&0.0565&130&9660\\
3C\,438&Y&0.290&22.6&48.72\\
\hline
\end{tabular}
\vskip 8pt
\begin{minipage}{\linewidth}
Redshifts and 178-MHz flux densities for these sources come from Laing
\etal\ (1983), updated to the more recent values given at
http://3crr.extragalactic.info/ where appropriate, with the flux densities
being given on the Baars \etal\ (1977) scale, except for 3C\,405,
where the flux density is the Baars \etal\ value.
\end{minipage}
\end{table}

The five new sources were then observed with BIMA\footnote{The BIMA array
was operated by the Berkeley Illinois Maryland Association under
funding from the U.S. National Science Foundation.}. Observational
details are given for all eight of the FRIIs discussed in this paper
in Table \ref{obs}. Our strategy in each case was to observe with the
BIMA configurations required to sample fully the largest-scale
structure seen in the source at lower frequencies. We observed the
largest sources with the BIMA D-array; the smallest source, 3C\,295,
only required B-array for full sampling, but we also used A-array to
resolve the source. In all cases, the correlator was configured to
give two 800-MHz bands around 86 GHz; for 3C\,295, these were centred
on 83.15 and 86.60 GHz, while for the other newly observed sources
they were at 86.00 and 89.44 GHz.

The new data were calibrated in {\sc miriad} (Sault, Teuben \& Wright 1995). 
Amplitudes were mostly calibrated by using observations of Mars and Uranus to
bootstrap a nearby bright point source, which was then used as a phase
calibrator. For the observations of 3C\,295, we used 3C\,273
as a flux calibrator, basing its flux density on monitoring
observations, which gave a flux density at our observing frequency of
11 Jy; at A configuration we made frequent observations of a nearby
bright quasar, J1419+543, to track short-timescale atmospheric phase
variations. The similar strategy adopted for 3C\,123 is discussed in
detail in Looney \& Hardcastle (2000). The error in the flux
density calibration is likely to be of the order of 10 per cent.

\begin{table}
\caption{Details of the BIMA observations of the
surface-brightness-selected sample}
\label{obs}
\begin{tabular}{lrlrr}
\hline
Source&Array&Date&Duration&Frequency\\
&&&(hours)&(GHz)\\
\hline
3C\,20&B&1999 Oct 30&7.7&84.9\\
&C&1999 Oct 11&6.4\\[2pt]
3C\,123&C&1996 Nov 02&9.9&107.8\\
&C&1996 Nov 06&8.8\\
&B&1997 Feb 15&11.2\\
&A&1996 Nov 27&10.1\\[2pt]
3C\,220.1&B&2001 Feb 01&7.6&87.7\\
&C&2000 Nov 04&3.7\\
&C&2000 Dec 06&4.2\\
&D&2000 Oct 12&5.5\\[2pt]
3C\,295&A&2000 Jan 26&10.6&84.9\\
&A&2000 Jan 31&8.6\\
&A&2000 Feb 01&4.7\\
&A&2000 Feb 02&10.3\\
&A&2000 Feb 05&9.1\\
&B&2000 Feb 23&5.7\\[2pt]
3C\,388&B&2001 Jan 21&8.6&87.7\\
&C&2000 Nov 05&8.0\\
&D&2000 Oct 05&6.0\\
&D&2000 Oct 17&3.2\\[2pt]
3C\,401&B&2001 Feb 14&8.1&87.7\\
&C&2000 Nov 20&8.0\\[2pt]
3C\,405&B&1997 Apr 26&8.6&87.7\\
&C&1996 Mar 31&6\\[2pt]
3C\,438&B&2001 Feb 27&6.0&87.7\\
&C&2000 Oct 26&7.4\\
&C&2000 Nov 28&6.3\\
\hline
\end{tabular}
\vskip 8pt
\begin{minipage}{\linewidth}
Only observations that were used in the analysis are listed.
Durations are for the entire run, including calibration observations;
this indicates the coverage of the $uv$ plane obtained. Typical
on-source integration times are 60 per cent of the run time for C and
D configurations, 50 per cent for B configuration, and 30 per cent for
A configuration. The frequency listed is the effective
frequency of the final images, the mean of the central frequencies of
the two BIMA bands used in each case.
\end{minipage}
\end{table}

\begin{figure*}
\epsfxsize 8.8cm
\epsfbox{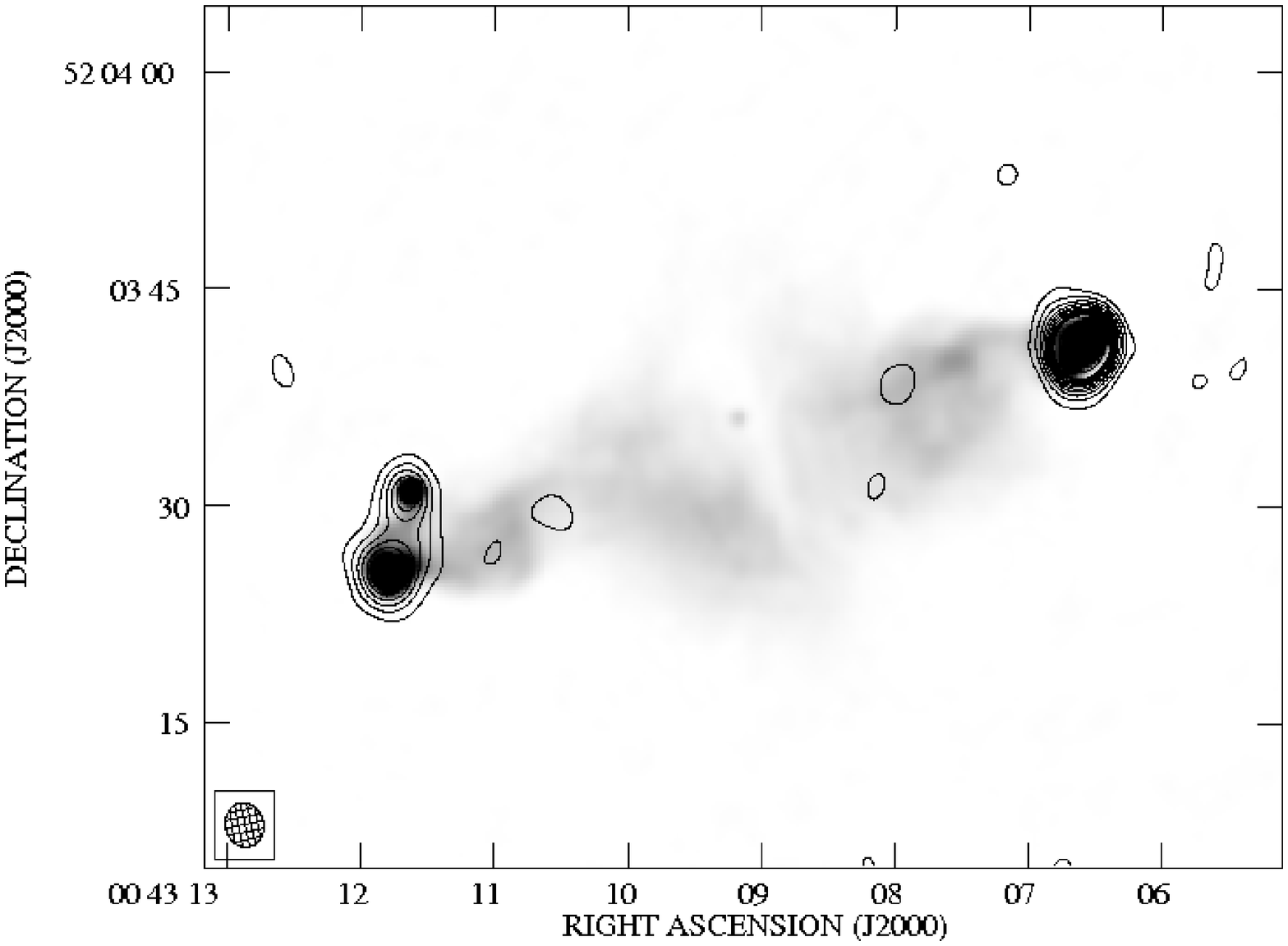}
\epsfxsize 8.8cm
\epsfbox{3C20.LARGE.PS}
\caption{BIMA images of 3C\,20. Left: contours of a $3.1 \times
  2.6$-arcsec resolution map (beam position angle $11.5\degr$) are
  superposed on a greyscale 1.1-arcsec 8.4-GHz VLA image (Hardcastle
  \etal\ 1997). Right: a $6.6 \times 5.9$-arcsec resolution BIMA map
  (beam position angle $19.7\degr$).
  Contours are at $-2, -1, 1, 2, 3, \dots, 10, 15, \dots 50, 60, \dots
  100$ times the $3\sigma$ level, which is 2.9 mJy beam$^{-1}$ for the
  high-resolution map and 3.4 mJy beam$^{-1}$ for the low-resolution
  one.}
\label{3C20}
\end{figure*}

\begin{figure*}
\epsfxsize 12cm
\epsfbox{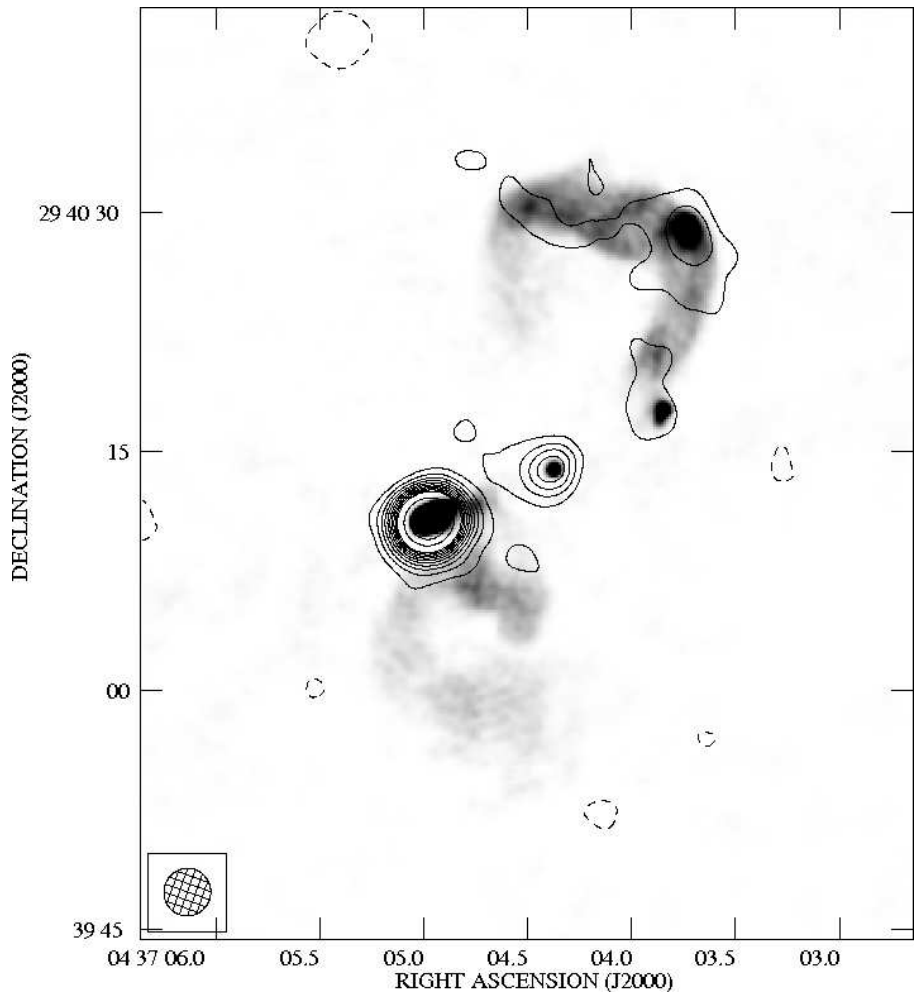}
\caption{BIMA image of 3C\,123. Contours of a $3.0 \times 2.9$-arcsec
  resolution map (beam position angle $-19.6\degr$) are superposed on
  a greyscale $0.6$-arcsec 8.4-GHz VLA image (Hardcastle
  \etal\ 1997). Contours as for Fig.\ \ref{3C20}: the $3\sigma$ level
  is at 9.0 mJy beam$^{-1}$.}
\label{3C123}

\end{figure*}

\begin{figure*}
\epsfxsize 12cm
\epsfbox{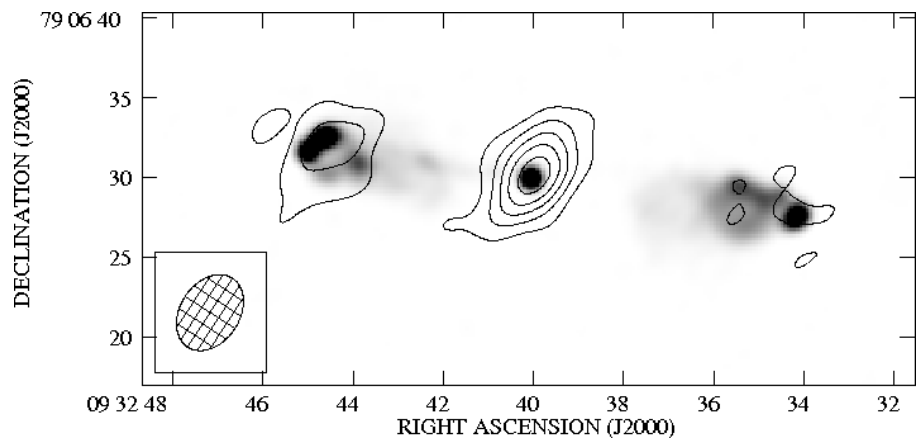}
\caption{BIMA image of 3C\,220.1. Contours of a $5.2 \times
3.7$-arcsec resolution map (beam position angle $-33.2\degr$) are superposed on a greyscale 0.6-arcsec
8.4-GHz VLA image (Mullin \etal\ 2006). Contours as for Fig.\ \ref{3C20}: the
$3\sigma$ level is at 2.6 mJy beam$^{-1}$.}
\label{3C220.1}
\end{figure*}

\begin{figure*}
\epsfxsize 12cm
\epsfbox{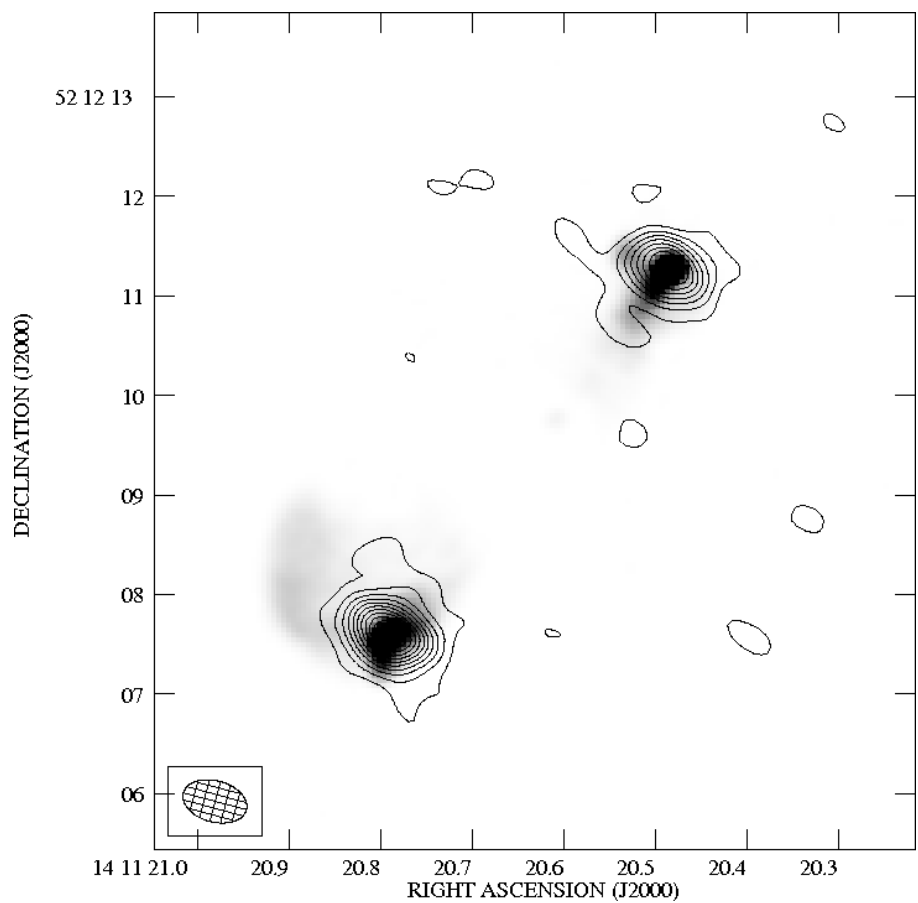}
\caption{BIMA image of 3C\,295. Contours of a $0.65 \times
  0.41$-arcsec resolution map (beam position angle $74.9\degr$) are superposed on a greyscale $0.18
  \times 0.17$-arcsec 8.5-GHz VLA image (beam position angle
  $13.8\degr$; Gilbert \etal\ 2004).
  Contours as for Fig.\ \ref{3C20}: the $3\sigma$ level is at 5.3 mJy
  beam$^{-1}$.}
\label{3C295}
\end{figure*}

\begin{figure*}
\epsfxsize 12cm
\epsfbox{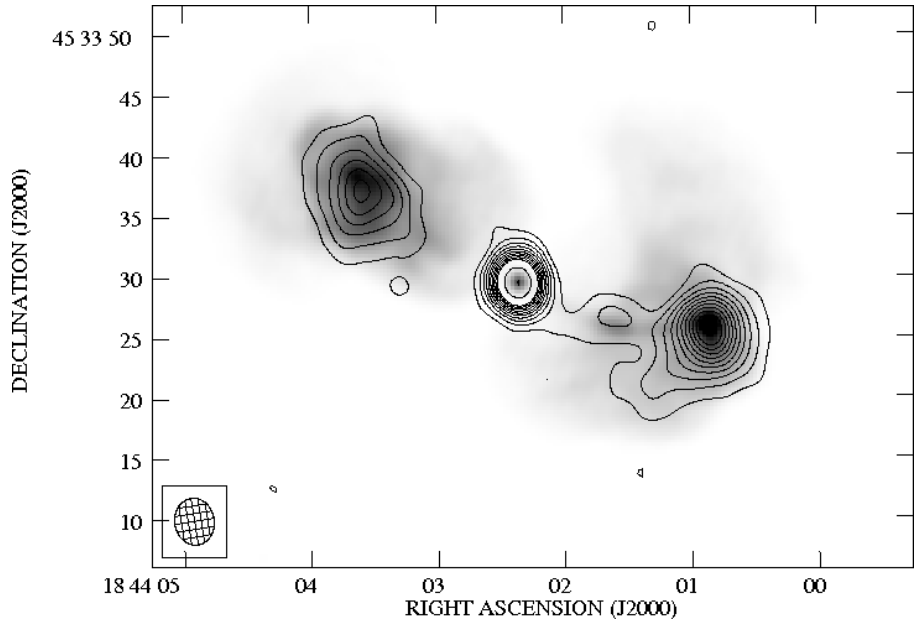}
\caption{BIMA image of 3C\,388. Contours of a $3.88 \times
3.27$-arcsec resolution map (beam position angle $8.9\degr$) are superposed on a greyscale
$1.32$-arcsec 1.4-GHz VLA image (Roettiger \etal\ 1994). Contours as
for Fig.\ \ref{3C20}: the $3\sigma$ level is at 1.6 mJy beam$^{-1}$.}
\label{3C388}
\end{figure*}

\begin{figure*}
\epsfxsize 10cm
\epsfbox{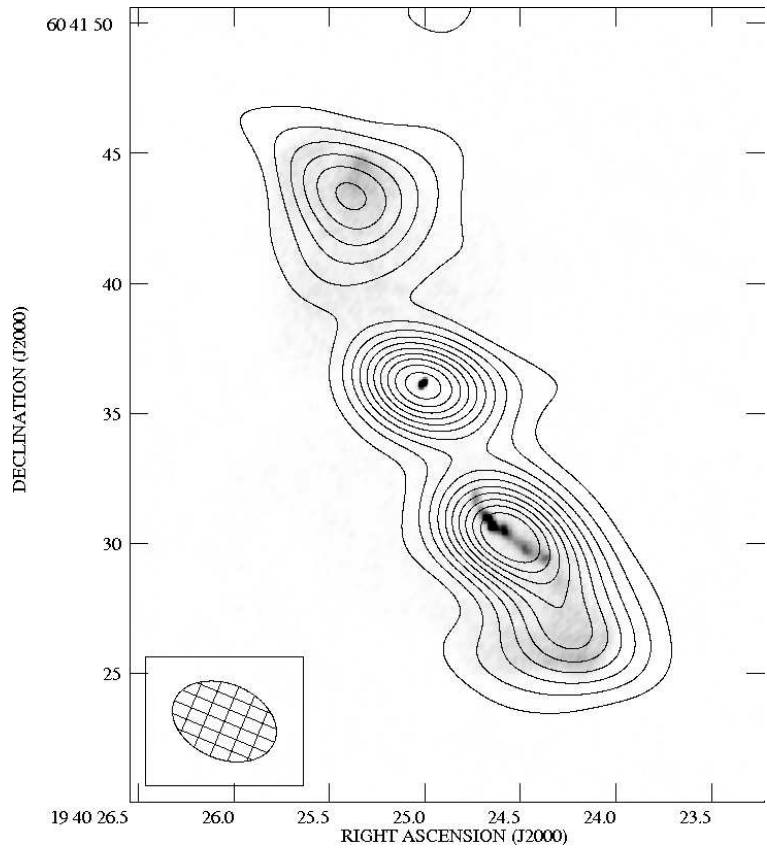}
\caption{BIMA image of 3C\,401. Contours of a $4.16 \times
2.89$-arcsec resolution map (beam position angle $67.4\degr$) are superposed on a greyscale $0.25 \times
0.17$-arcsec
8.4-GHz VLA image (beam position angle $-36.0\degr$; Hardcastle \etal\ 1997). Contours as for Fig.\ \ref{3C20}: the
$3\sigma$ level is at 1.2 mJy beam$^{-1}$.}
\label{3C401}
\end{figure*}

\begin{figure*}
\epsfxsize 12cm
\epsfbox{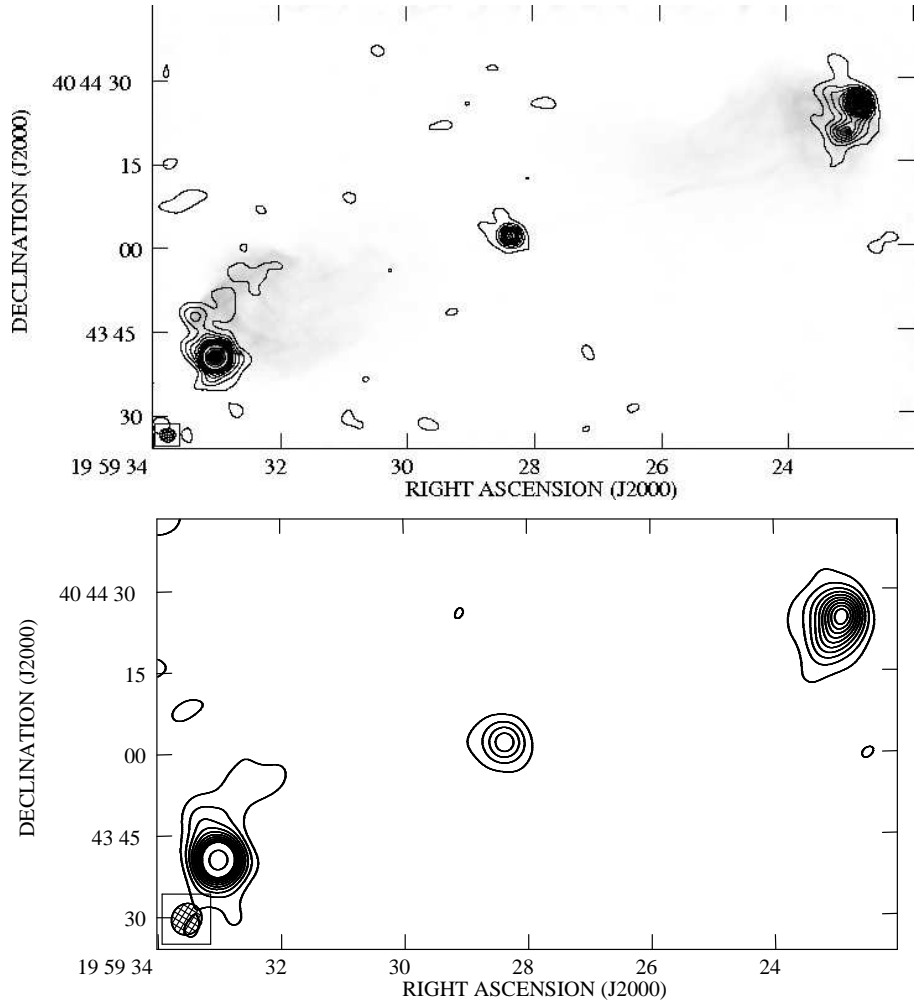}
\epsfxsize 12cm
\epsfbox{3C405.LOW.PS}
\caption{BIMA image of Cygnus A (3C\,405). Top: contours of a $2.83
  \times 2.44$-arcsec resolution map (beam position angle $60.8\degr$)
  are superposed on a greyscale 0.4-arcsec 4.9-GHz VLA image (Dreher,
  Perley \& Cowan 1984). Bottom: a $5.93 \times 5.43$-arcsec
  resolution map (beam position angle $-31.7\degr$). Contours as for
  Fig.\ \ref{3C20}: the $3\sigma$ level for the first map is 60 mJy
  beam$^{-1}$ and for the second is 160 mJy beam$^{-1}$, including in
  both cases the effects of positive bias from the maximum entropy
  routine used for the deconvolution.}
\label{3C405}
\end{figure*}

\begin{figure*}
\epsfxsize 12cm
\epsfbox{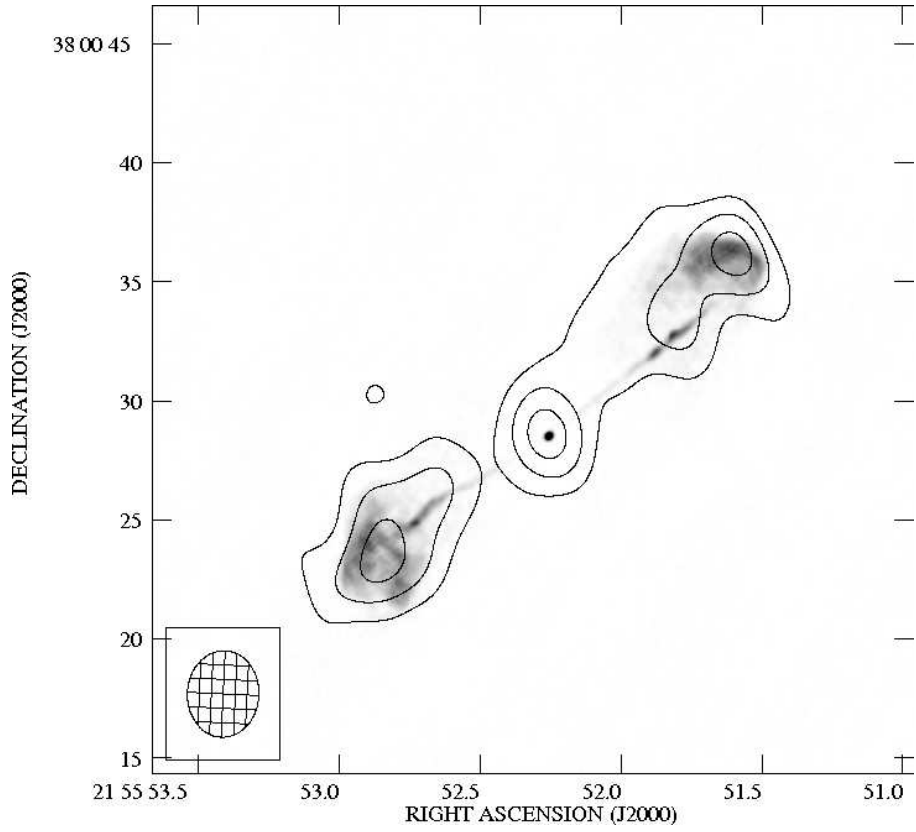}
\caption{BIMA image of 3C\,438. Contours of a $3.63 \times
3.00$-arcsec resolution map (beam position angle $-2.7\degr$) are superposed on a greyscale $0.25 \times
0.21$-arcsec 8.4-GHz VLA image (beam position angle $-47.8\degr$; Hardcastle \etal\ 1997). Contours as
for Fig.\ \ref{3C20}: the $3\sigma$ level is at 1.8 mJy beam$^{-1}$.}
\label{3C438}
\end{figure*}

The datasets at the various different BIMA configurations were then
combined, and final images were made in {\sc aips} using the {\sc
imagr} task, except for the observations of Cygnus A, where the joint
deconvolution algorithms provided by {\sc miriad} were used to image
each sideband, and the final images were combined in {\sc aips}. For a
typical source, where the longest baselines were provided by the B
configuration, the full resolution of the combined datasets was about
3 arcsec, and the off-source noise at that resolution was between 0.5
and 1 mJy beam$^{-1}$. In Figs \ref{3C20} to \ref{3C438} we show
images of all these sources.

\begin{table}
\caption{The additional sample of FRII radio sources}
\label{addsamp}
\begin{tabular}{llrrr}
\hline
Source&Newly&$z$&LAS&178-MHz flux\\
&observed?&&(arcsec)&density (Jy)\\
\hline
3C\,263&N&0.6563&51&16.6\\
3C\,330&N&0.5490&60&30.3\\
3C\,351&N&0.371&74&14.9\\
\hline
\end{tabular}
\vskip 8pt
\begin{minipage}{\linewidth}
Notes as for Table \ref{sample}.
\end{minipage}
\end{table}

\begin{table}
\caption{Details of the BIMA observations of the
additional sample}
\label{addsobs}
\begin{tabular}{lrlrr}
\hline
Source&Array&Date&Duration&Frequency\\
&&&(hours)&(GHz)\\
\hline
3C263&B&2002 Feb 03&8.0&84.9\\
3C330&B&2002 Feb 08&4.8&84.9\\
3C351&B&2002 Feb 10&7.5&84.9\\
\hline
\end{tabular}
\vskip 8pt
\begin{minipage}{\linewidth}
Notes as for Table \ref{obs}.
\end{minipage}
\end{table}

\begin{figure*}
\epsfxsize 14cm
\epsfbox{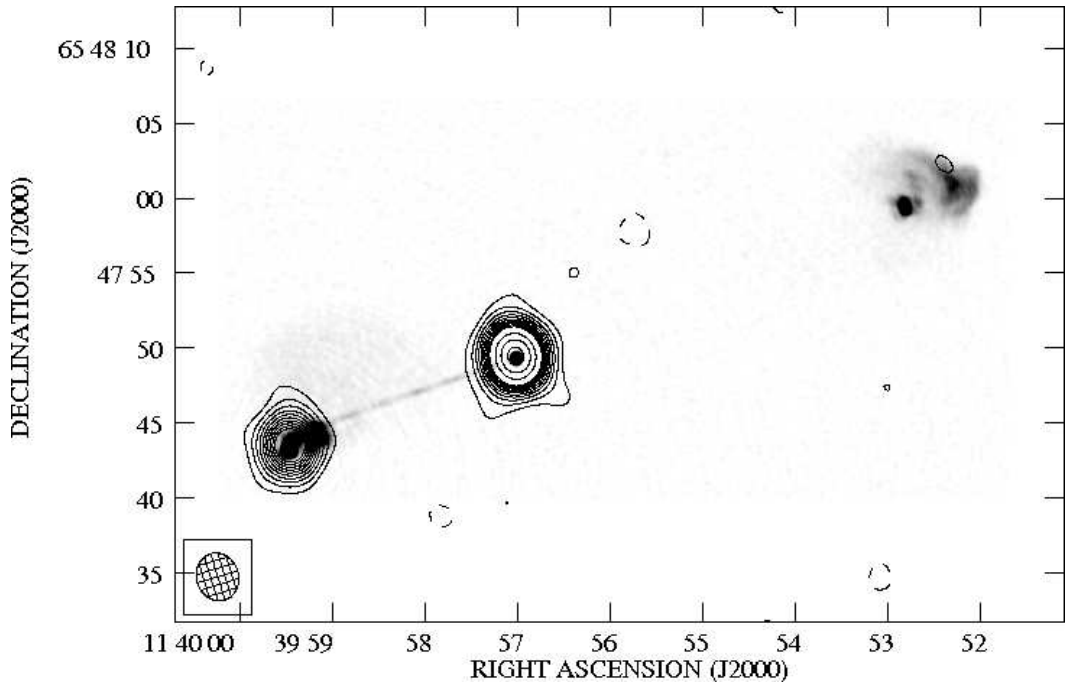}
\caption{BIMA image of 3C\,263. Contours of a $3.22 \times
  2.79$-arcsec resolution map (beam position angle $17.0\degr$) made with B-configuration data only are
  superposed on a greyscale $0.36$-arcsec 4.9-GHz VLA image (Bridle
  \etal\ 1994). Contours as for Fig.\ \ref{3C20}: the $3\sigma$ level
  is at 1.1 mJy beam$^{-1}$.}
\label{3C263}
\end{figure*}

\begin{figure*}
\epsfxsize 14cm
\epsfbox{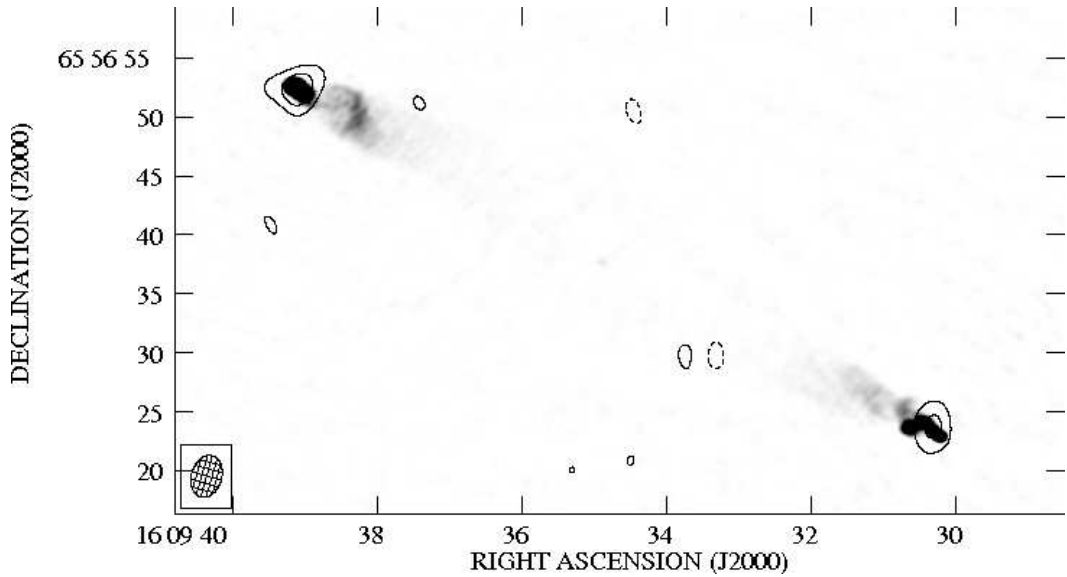}
\caption{BIMA image of 3C\,330. Contours of a $3.61 \times
  2.56$-arcsec resolution map (beam position angle $29.1\degr$) made with B-configuration data only are superposed on a greyscale $0.57
  \times 0.48$-arcsec 8.4-GHz VLA image (beam position angle
  $76.7\degr$; Gilbert \etal\ 2004).
  Contours as for Fig.\ \ref{3C20}: the $3\sigma$ level is at 2.7 mJy
  beam$^{-1}$.}
\label{3C330}
\end{figure*}

\begin{figure*}
\epsfxsize 10cm
\epsfbox{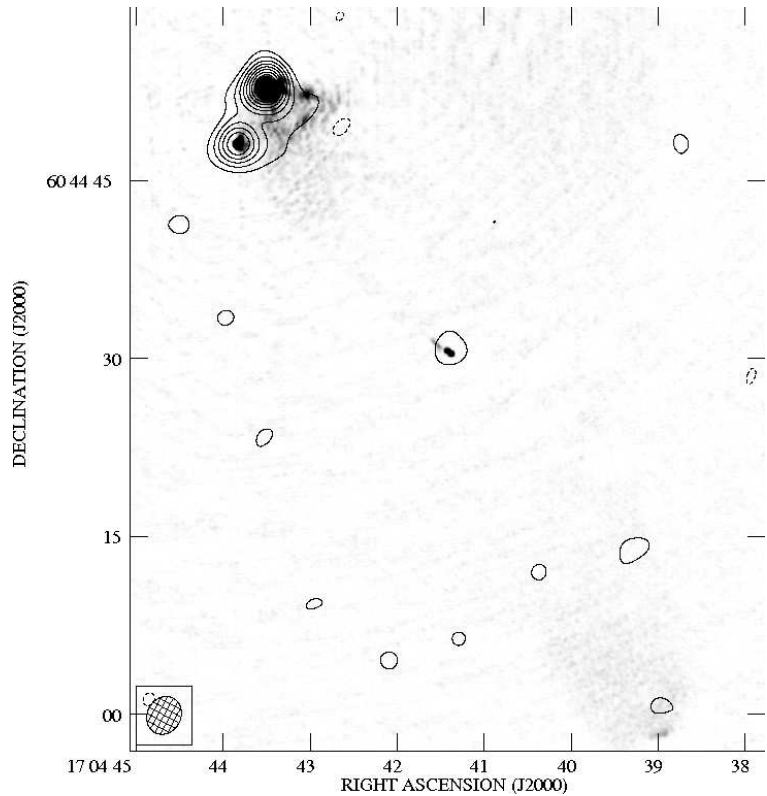}
\caption{BIMA image of 3C\,351. Contours of a $3.28 \times
  2.83$-arcsec resolution map (beam position angle $-28.0\degr$) made
  with B-configuration data only are superposed on a greyscale
  $0.3$-arcsec 8.4-GHz VLA image (Gilbert \etal\ 2004). Contours as
  for Fig.\ \ref{3C20}: the $3\sigma$ level is at 0.94 mJy
  beam$^{-1}$.}
\label{3C351}
\end{figure*}

Hardcastle \etal\ (2002) discuss B-array BIMA observations of three
further FRII sources, one radio galaxy (3C\,330) and two quasars
(3C\,263 and 3C\,351). These objects
were not selected for high lobe surface brightness, but for hotspot
brightness: moreover, the B-array observations are not optimized for
lobe detections, and so can only tell us about compact source
components. We make use of these observations to increase our sample
of core and hotspot measurements. Details of this additional sample
are given in Table \ref{addsamp} and the observing parameters are
given in Table \ref{addsobs}. The two observing bands used were at
83.2 and 86.6 GHz. Images of these three sources are shown in Figs
\ref{3C263} to \ref{3C351}.

\section{Results: components of radio sources}

As Figs \ref{3C20}--\ref{3C351} show, BIMA detects emission from the
cores, jets, hotspots and (in some cases) the lobes of the sources in
the samples. In this section we discuss the trends seen for each of these
components. Throughout we refer to the BIMA observing frequency for
simplicity as 90
GHz, though the correct effective frequency for each source is used
when calculating two-point spectral indices.

\subsection{Cores}

Eight of the eleven sources in the combined sample show 90-GHz core
components. In Table \ref{cores} we tabulate flux densities for the
cores of our sample at centimetre and millimetre wavelengths.

These results show that it is normal for the arcsecond-scale cores of
both radio galaxies and quasars to have fairly flat spectra
($\alpha_{\rm cm}^{\rm mm}<0.5$) over this range of radio frequencies.
In most cases the mm-wave flux density is less than the highest value
measured at lower wavelengths, which suggests that core spectra
typically peak at intermediate frequencies of tens of GHz. However, there is no
sign of a cutoff in the spectra of cores by 90 GHz. Similar results
have been obtained for other individual low-redshift radio galaxies;
for example, Das \etal\ (2005) detect a 90-GHz core at a level of
$\sim 15$ mJy in the nearby ($z=0.011$) FRI radio galaxy NGC 3801,
which has a 5-GHz core flux density of 3.5 mJy (Croston \etal\ 2007)
while at 8.4 GHz archival VLA observations give a core flux density of
5.9 mJy, implying a substantially inverted spectrum. Studies of
core-dominated objects such as blazars at high frequencies (e.g. Bloom
\etal\ 1994) show that flat radio-mm spectra are also typical in those
objects. One might expect that our results in this paper would imply
that high-redshift radio galaxies would show flat core radio spectra
at lower observing frequencies; interestingly, though, this is not
consistent with the results of Athreya \etal\ (1997), who found
predominantly steep ($\alpha>0.5$) spectra at rest-frame frequencies
as low as 20--30 GHz in a sample of $z>2$ radio galaxies. This may be
evidence for a true redshift and/or luminosity dependence of the
rest-frame behaviour of radio cores, but we emphasise that our sample
(by virtue of its selection) is by far from being representative of
radio galaxies and lobe-dominated quasars as a whole; observations of
a complete sample are required to draw definitive conclusions.

None of the sources for which we have multi-epoch BIMA data shows
evidence for strong mm-wave core variability on timescales of months;
the core flux densities in individual observations are consistent
within the errors. Two of our sources, 3C\,123 (Looney \& Hardcastle
2000) and Cygnus A (Wright \& Sault 1995) show some evidence for
variability on longer timescales when the BIMA core flux density is
compared with that measured with other instruments. None of our
targets is known to have a strongly variable core at longer
wavelengths.
 
\begin{table}
\caption{Arcsecond-scale core flux densities for the samples at centimetre and
millimetre wavelengths}
\label{cores}
\begin{tabular}{lrrrrr}
\hline
Source&\multicolumn{5}{c}{Core flux density (mJy)}\\
&1.4 GHz&5 GHz&8.4 GHz&15 GHz&BIMA\\
\hline
3C\,20&&2.6$^a$&3.3$^b$&&$<3$$^c$\\
3C\,123&64$^d$&93$^d$&90$^d$&96$^d$&42$^d$\\
3C\,220.1&&25$^e$&27$^n$&&15$^g$\\
3C\,295&&&4.6$^l$&&$<5$$^g$\\
3C\,388&50$^h$&62$^e$&56$^m$&&33$^g$\\
3C\,401&17$^f$&32$^e$&28$^f$&&17$^g$\\
3C\,405&&776$^i$&&&887$^g$\\
3C\,438&7$^f$&&16$^f$&&6$^g$\\
\hline
3C\,263&126$^j$&157$^k$&121$^j$&141$^j$&108$^g$\\
3C\,330&$<5^j$&0.74$^a$&0.63$^l$&$<6^j$&$<5^g$\\
3C\,351&$<36^j$&6.5$^k$&6.5$^j$&&2$^g$\\
\hline
\end{tabular}
\begin{minipage}{\linewidth}
For simplicity, we do not tabulate the exact frequency of each
measurement. Upper limits are $3\sigma$ limits based on the off-source
noise, except where otherwise stated below. References are as follows:
(a) Fernini, Burns \& Perley (1997) (b) Hardcastle \etal\ (1998) (c)
Hardcastle \& Looney (2001) (d) Looney \& Hardcastle (2000) (e)
Giovannini \etal\ (1988) (f)
Measured from maps used by Treichel \etal\ (2001) (g) This paper (h)
Measured from map obtained from the 3CRR Atlas
(http://www.jb.man.ac.uk/atlas/) (i) Measured from the map of Perley,
Dreher \& Cowan (1984) (j) Measured from a map used by Hardcastle
\etal\ (2002) (k) Bridle \etal\ (1994) (l) From maps of Gilbert \etal\
(2004) (m) Measured from the map used by Kraft \etal\ (2006) (n)
Mullin \etal\ (2006). For
3C\,351, only the southern component of the two described by Bridle
\etal\ (1994) is included in the 5- and 8-GHz flux density measurements, while
the 1.4-GHz measurement is considered an upper limit because the
resolution is too low to separate the two. Note that this table takes
no account of possible core variability at any frequency.
\end{minipage}
\end{table}

\subsection{Jets}

\begin{table*}
\caption{Jet flux densities and two-point spectral indices for sources
with detected jets}
\label{jets}
\begin{tabular}{lrrrrr}
\hline
Jet&\multicolumn{3}{c}{Jet flux density (mJy)}&\multicolumn{2}{c}{Spectral index}\\
&1.4 GHz&8.4 GHz&BIMA&$\alpha_{1.4}^{8.4}$&$\alpha_{8.4}^{88}$\\
\hline
3C\,388 W&$63.8 \pm 1.9$&$25.9 \pm 0.8$&$5.7 \pm 1.0$&$0.50 \pm 0.02$&$0.65 \pm 0.07$\\
3C\,401 S&$333 \pm 10$&$129 \pm 4$&$25.5\pm2.8$&$0.53 \pm 0.02$&$0.69 \pm 0.05$\\
3C\,438 N&$111 \pm 3$&$44 \pm 1$&$6.2 \pm 0.9$&$0.52 \pm 0.02$&$0.83
\pm 0.06$\\
3C\,438 S&$99 \pm 3$&$30 \pm 1$&$3.3 \pm 0.6$&$0.62 \pm 0.02$&$0.98
\pm 0.07$\\
\hline
\end{tabular}
\end{table*}

Jet-related emission is clearly detected in three of the newly
observed sources, 3C\,388, 3C\,401, and 3C\,438. This is the first
time that FRII jets have been detected at these high radio frequencies. No
jet-related emission is detected from five other sources (3C\,20,
3C\,220.1, 3C\,263, 3C\,351 and 3C\,405) that have known
centimetre-wave jet detections, but the jets in these objects are
considerably fainter at lower frequencies. The three detected sources
are all part of the class of low-excitation radio galaxies with
prominent jets and diffuse hotspots, possibly lying in rich
environments, identified by Hardcastle \etal\ (1997).

To investigate the spectra of the jets we used the 1.4 and 8.4-GHz
radio maps listed in Table \ref{cores} and the full-resolution BIMA
datasets. We defined regions on the BIMA maps that include all the
apparently jet-related flux density, and then extracted all the apparently
jet-related flux density from the (typically much higher resolution)
VLA/MERLIN maps. For the GHz-frequency maps the dominant source of
(non-systematic) error is the absolute flux density calibration of the images,
which we estimate conservatively at 3 per cent. Systematic errors are
dominated by the requirement to subtract a background due to the lobes
-- we did this using identical offset regions adjacent to the jet. For
the BIMA images, the thermal noise in the extraction region and the
overall estimated flux density calibration error are added in quadrature to
give a flux density error estimate. Results are tabulated in Table
\ref{jets}.

In all three of the jet sources we see that the two-point spectral
indices between low and high frequencies are significantly
inconsistent, in the sense that the spectrum is steeper between cm and
mm wavelengths. Thus the spectrum cannot be modelled with a single power
law between 1.4 and 90 GHz -- a spectral break is required. As the
jets in these regions are knotty and complex, a single simple spectral
model is probably not appropriate anyway, but it does suggest that
these jets are not similar in this respect to FRI jets, which in spite
of knotty and complex structure can have a fairly flat spectrum out to
the optical and beyond. Chiaberge \etal\ (2005) report a detection of
the 3C\,401 jet in {\it Hubble Space Telescope} near-infrared images,
implying a flux density of 6.0 $\mu$Jy at 190 THz, which would
correspond to a spectral index between the mm and infrared of $\sim
0.74$: thus it seems that the spectrum of the jet breaks at
frequencies between 10 and 100 GHz and then remains roughly constant
over many decades in frequency. Nothing is known about the
high-frequency behaviour of the other two jets, although they are not
detected in {\it Chandra} images.

The brightest non-detected jet in the sample is that of 3C\,263
(Bridle \etal\ 1994), with a radio flux density around 13 mJy: the
$3\sigma$ upper limit on the 90-GHz flux density from the jet region
is 6 mJy, so no interesting limit can be placed on the jet spectral index.

\subsection{Hotspots}

Almost all of our targets have clear detections of the region at or
around the hotspots at 90 GHz. Only the faintest cm-wave hotspots
(3C\,263 W and 3C\,351 S) are undetected in these observations. Since
our observations sample a wide range of hotspot structures, we can
conclude that it is plausible that few or no FRII hotspots -- at least
in large, powerful FRIIs -- have
spectra that cut off below the mm band. This is of course unsurprising
in view of the detections of optical, infrared and sometimes X-ray
synchrotron radiation from hotspots (Meisenheimer, Yates \& R\"oser 1997;
Hardcastle \etal\ 2004) and the fact that the mm-wave observations
probe electron energies only a factor 3 above the highest-frequency
cm-wave observations. The high-energy cutoff for particle acceleration
is set by the balance between acceleration and loss, and it seems that
the very high energy loss rates (implying high photon and/or magnetic
field energy densities) required to push that cutoff down to a level
where it appears at cm or mm wavelengths are not found in the type of
objects we have studied here, though they may be present in some compact
steep-spectrum sources (e.g. 3C\,196, Hardcastle 2001).

The spectra of the hotspots of many of our sources have been discussed
in detail elsewhere [3C\,405, Wright \etal\ (1997); 3C\,123, Looney \&
Hardcastle (2000); 3C\,20, Hardcastle \& Looney (2001); 3C\,263,
3C\,330 and 3C\,351, Hardcastle \etal\ (2002)]. In general these
results have confirmed what has been known for some time from other
work (e.g. Meisenheimer \etal\ 1989): some sources' hotspots have
spectra that are flat up to optical or even X-ray frequencies, while
others exhibit a clear spectral steepening either between cm and mm
wavelengths, as was the case for 3C\,123, or between the mm-wave and
the near-infrared. In Table \ref{hotspots} we provide a compilation of
radio and mm-wave flux densities for our sample. Since in general the
cm-wavelength maps have higher resolution than the mm-wavelength ones,
we adopt the approach (discussed in our earlier papers) of quoting
only the flux density of the most compact components accessible to us at radio
wavelengths, determined by direct integration with background
subtraction where these are well resolved and fitting a Gaussian and
baseline when they are not. At 90 GHz we quote flux densities measured
by direct integration of the apparently hotspot-related region of the
map. This is the correct thing to do if we believe that all the flux
density at 90 GHz is due to the hotspot; if there is in fact excess
(lobe-related) 90-GHz emission in our integration region, we will tend
to overestimate the 90-GHz hotspot flux density, but this seems inevitable
without higher-resolution mm-wave observations. We have chosen not to
tabulate hotspot flux densities for the three sources (3C\,388,
3C\,401 and 3C\,438) with diffuse hotspots at cm wavelengths, since for
these sources we cannot distinguish adequately between hotspots and
lobe emission (see section \ref{lobe-section}, below).

\begin{figure}
\epsfxsize 8.5cm
\epsfbox{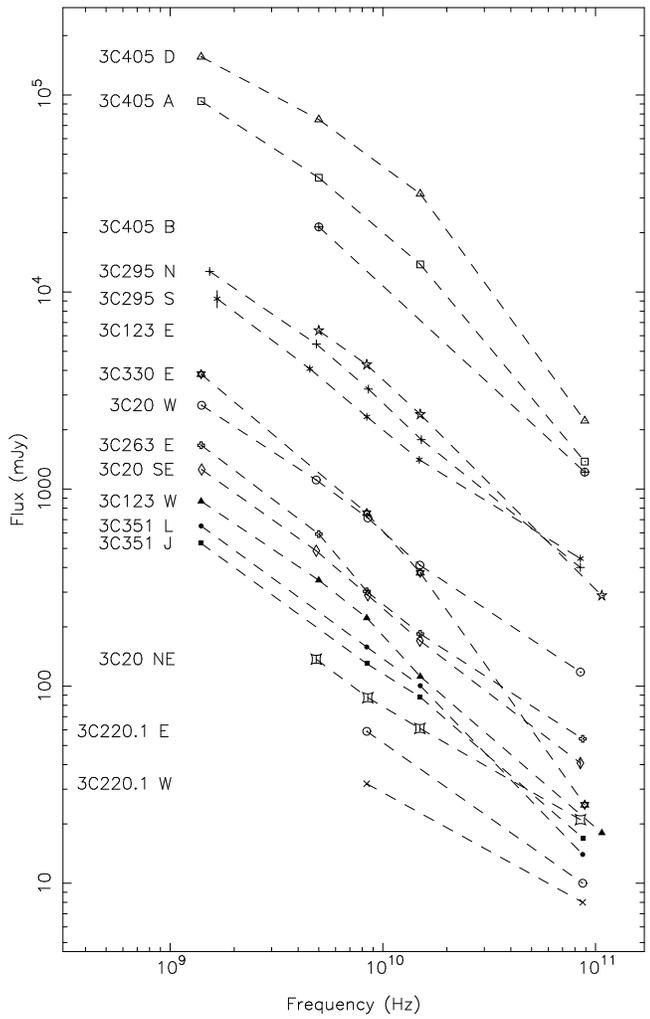}
\caption{Spectra of the BIMA-detected hotspots listed in Table
  \ref{hotspots}. Some flux densities have been scaled for clarity. 3C\,405 D
  has been increased by a factor 1.5; 3C\,405 B, 2.0; 3C\,295 N, 5.0;
  3C\,295 S, 4.0; 3C\,351 L, 0.5.}
\label{hsfig}
\end{figure}

The contents of Table \ref{hotspots} are plotted in Fig. \ref{hsfig}.
This illustrates the wide range of cm-mm spectral indices found in the
data ($\alpha$ from $\sim 0.6$ to $\sim 1.5$). One trend that is
apparent is that in several cases, {\it within a given source},
brighter hotspots have steeper cm-to-mm spectra ($\alpha_{\rm cm}^{\rm
mm}$). In several of these objects the brighter hotspots are large
secondaries (as in 3C\,20, 3C\,351 and 3C\,405) and so this may be
connected to the observation that large diffuse secondary hotspots in
multiple-hotspot sources tend not to show X-ray synchrotron radiation
(Hardcastle, Croston \& Kraft 2007); as discussed in previous work
(e.g. Hardcastle \& Looney 2001) there is little evidence for spectral
ageing effects in secondary hotspots at these frequencies. The data
also hint that more luminous hotspots have steeper $\alpha_{\rm
cm}^{\rm mm}$, but there are insufficient sources to carry out
meaningful statistical comparisons.

\begin{table*}
\caption{Hotspot flux densities for the samples at centimetre and
millimetre wavelengths}
\label{hotspots}
\begin{tabular}{llrrrrr}
\hline
Source&Hotspot&\multicolumn{5}{c}{Flux density (Jy)}\\
&&1.4 GHz&5 GHz&8.4 GHz&15 GHz&BIMA\\
\hline
3C\,20&W&2.661&1.112&0.714&0.413&0.118\\
&NE&&0.137&0.087&0.061&0.021\\
&SE&1.257&0.486&0.289&0.171&0.037\\[2pt]
3C\,123&E&&6.372&4.277&2.392&0.288\\
&W&0.867&0.345&0.222&0.112&0.018\\[2pt]
3C\,220.1&E&&&0.059&&0.010\\
&W&&&0.032&&0.008\\[2pt]
3C\,295&N&2.545&1.089&0.643&0.356&0.080\\
&S&0.231&0.102&0.582&0.352&0.111\\
3C\,405&A&93&38&&13.8&1.38\\
&B&&10.7&&&0.61\\
&D&104&50&&21&1.48\\
\hline
3C\,263&E&1.670&0.591&0.303&0.184&0.054\\
&W&&0.023&0.016&0.009&$<0.004$\\[2pt]
3C\,330&E&3.830&&0.755&0.376&0.025\\
&W&&&0.079$\dag$&&0.018\\[2pt]
3C\,351&N (J)&0.530&&0.130&0.088&0.017\\
&N(L)&1.30&&0.316&0.201&0.028\\
&S&$<0.017$&&0.003&0.0017&$<0.003$\\
\hline
\end{tabular}
\begin{minipage}{\linewidth}
Upper limits are $3\sigma$ limits based on the off-source noise.
References are as follows: 3C\,20, Hardcastle \& Looney (2001) (we
tabulate the `compact' component values for the NE hotspot but the
`region' values for the W hotspot); 3C\,123, Looney \& Hardcastle
(2000); 3C\,263, 3C\,330 \& 3C\,351, Hardcastle \etal\ (2002);
3C\,295, radio flux densities from Harris \etal\ (2000); 3C\,405
hotspots A and D, radio flux densities from Carilli \etal\ (1991); all
other values, this paper. For 3C\,405 we use the conventional notation
of Carilli \etal\ (1991) to label the hotspots; hotspot A and B are
respectively the secondary and primary hotspots in the W lobe and D
the brightest hotspot in the E lobe. Similarly for 3C\,351 we use the
notation of Bridle \etal\ (1994), in which J is the compact primary
hotspot and L is the secondary. \\ $\dag$ We take the opportunity
to correct the incorrectly transcribed flux density for 3C\,330 W that
was quoted by Hardcastle \etal\ (2002).
\end{minipage}
\end{table*}

\subsection{Lobes}
\label{lobe-section}

It is very clear that hotspots, cores and (where present) jets
dominate the high-frequency radio emission from these sources.
Nevertheless some sources show clear emission from lobe regions. These
are 3C\,20 and 3C\,123 (as described in our earlier papers), 3C\,405
(where a spur of emission leading N and W from the E hotspot is seen)
and 3C\,388 (where the W hotspot is extended to the SE). In all four
of these cases the `lobe' emission at mm wavelengths is well matched
morphologically to high-surface-brightness features close to the
hotspot at cm wavelengths. As noted above, extended emission in
3C\,388 E, 3C\,401 and 3C\,438 cannot unambiguously be associated with
either the lobes or the hotspots, but it is plausible that some or all
of this emission is also related to the lobes. No lobe-related
emission is seen in 3C\,220.1 or 3C\,295, though of course these are
the highest-redshift sources in our sample with observing frequencies
corresponding to $\ga 140$ GHz in the source frame.

\begin{figure*}
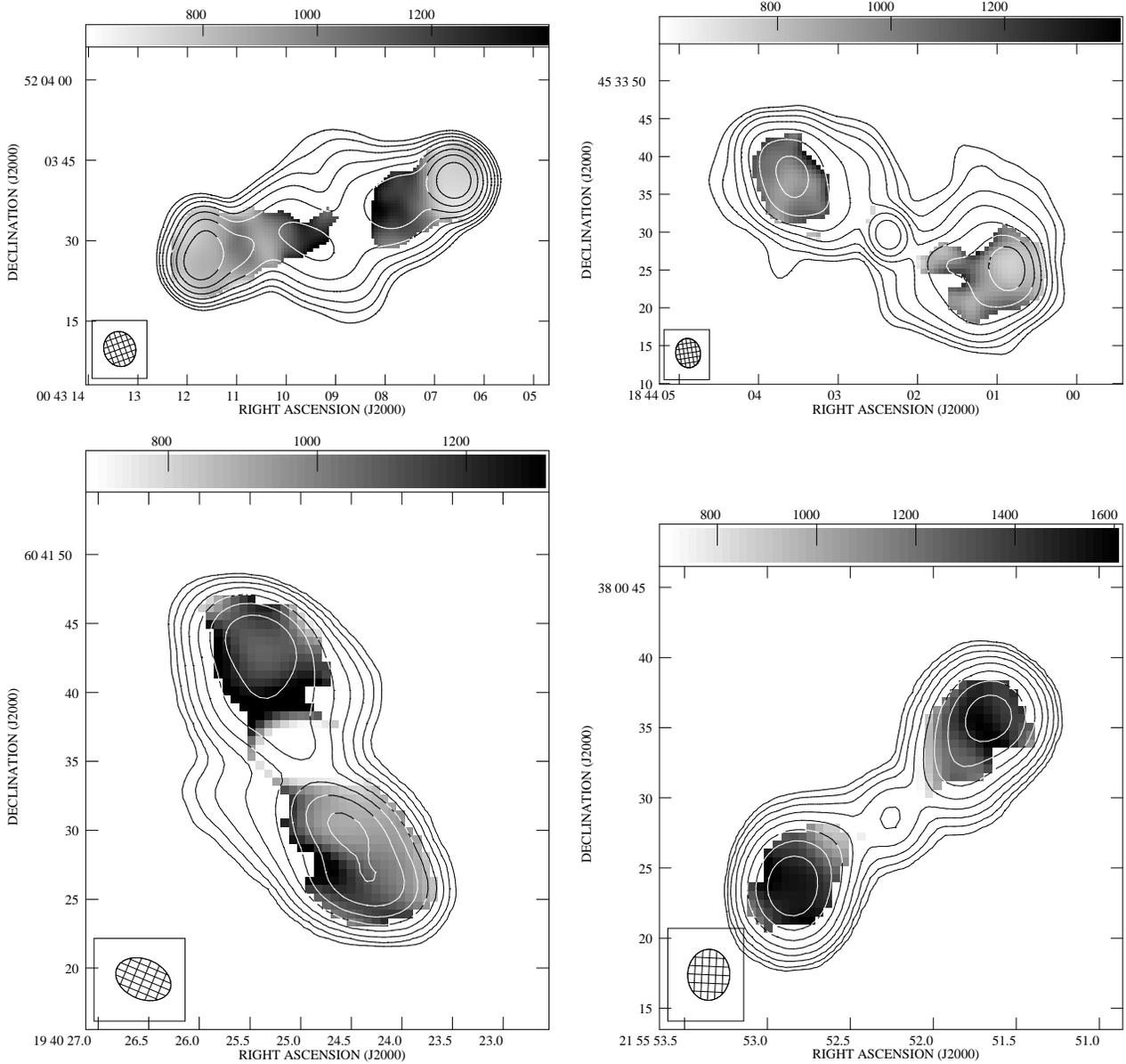

\hbox{\epsfxsize 8.5cm
\epsfbox{3C20.SPIX.PS}
\epsfxsize 8.5cm
\epsfbox{3C388.SPIX.PS}
}
\hbox{\epsfxsize 8.5cm
\epsfbox{3C401.SPIX.PS}
\epsfxsize 8.5cm
\epsfbox{3C438.SPIX.PS}
}
\caption{Maps of spectral index between 8.4 GHz and 90 GHz. Top left:
  3C\,20 at $6.6 \times 5.9$ arcsec resolution (beam position angle
  $19.7\degr$). Top right: 3C\,388 at $3.88 \times 3.27$ arcsec
  (position angle $8.8\degr$). Bottom left: 3C\,401 at $4.19 \times
  2.89$ arcsec (position angle $67.3\degr$). Bottom right: 3C\,438 at
  $3.63 \times 3.00$ arcsec (position angle $-2.8\degr$). Colour bars
  at the top of images are in units of $1000 \times
  \alpha_{8.4}^{90}$. Contours overlaid are from matched-resolution
  8.4-GHz maps in each case. These maps were made from the VLA $uv$
  data used by Hardcastle \etal\ (1997), Kraft \etal\ (2006),
  Hardcastle \etal\ (1997) and Treichel \etal\ (2001) respectively.
  White contours are used when the underlying spectral index map is
  dark.}
\label{spix}
\end{figure*}

In Fig.\ \ref{spix} we show maps of spectral index between 8.4 and 90
GHz for the four small sources with detected emission that might be
lobe-related. 3C\,20 (as already discussed in Hardcastle \& Looney
2001) and 3C\,388 show clear spectral steepening away from the
hotspots. In 3C\,401 and 3C\,438 the flattest-spectrum regions are the
jets (already discussed above) and there is no clear evidence for
resolved spectral steepening in the lobes. 

We concentrate initially on the sources with clear detections of
lobe-related emission at 90 GHz: 3C\,20, 3C\,123, 3C\,388 and 3C\,405.
The spectra of the lobes of the first two is discussed in our earlier
papers: we measured flux densities in matched regions defined using
the 90-GHz maps for 3C\,388 and 3C\,405. In all of these sources we
find that the spectral index between cm wavelengths (5 or 8 GHz) and
90 GHz is steeper in the extended emission than in the corresponding
hotspot, which is qualitatively consistent with expectations from
spectral ageing models. However, if there is any curvature in the
hotspot spectra (as there is in the spectra of the 3C\,123 and 3C\,405
hotspots) then a simple decrease in the magnetic field strength and
electron energy, as expected if plasma expands adiabatically on
leaving the hotspots, would also produce such a result. As discussed
in Section \ref{intro}, we showed in the case of the peculiar object
3C\,123 that the spectrum of the detected northern lobe region is
inconsistent with a Jaffe \& Perola (1973; hereafter J-P) spectral
ageing model, perhaps a result of particle reacceleration in the lobe.
By contrast the spectrum of the 3C\,20 E lobe is well fitted with a
J-P model provided that the `injection index' (the low-frequency
spectral index) is tuned to allow a good fit to the low-frequency
data. In 3C\,388, the only other source where we have multi-frequency
data on the lobes readily available, we find that the same is true --
a J-P model provides a good fit to the data from the W lobe provided
that the injection index is $\sim 0.7$. Our data are not sensitive
enough to take the next step and produce maps of spectral age as a
function of position to test whether the spectral ages required are
physically reasonable and/or consistent with what is found at lower
frequencies.

For the sources with undetected lobes (3C\,220.1 and 3C\,295) we can
only set lower limits on the spectral indices of any undetected lobe
emission of $\alpha_{8.4}^{90}$ (lab frame) $> 1.2$ for both sources. This is
consistent with the measured spectral indices in the steepest-spectrum
regions of other sources (Fig.\ \ref{spix}) and
so there is little evidence for a difference in physical behaviour in
these two objects.

\section{Discussion}

\subsection{Summary: radio galaxies at 90 GHz}

Our observations have now detected at 90 GHz the synchrotron emission
from arcsec-scale components of FRII radio galaxies seen at lower
frequencies: cores, jets, hotspots and lobes. We can summarize our
results and their consequences as follows:

\begin{itemize}
\item Cores are flat-spectrum up to 90 GHz in most cases, implying
  that synchrotron self-absorption continues to be important in the cores well
  above cm wavelengths.

\item Bright jets are detected and show curved or broken power-law
  spectra, which is interesting given that the mechanism of particle
  acceleration in FRII jets is not well known at present. Particle
  acceleration in jets is required for the jets to be visible (unless
  they are simply boundary-layer effects) but is essentially ignored
  in most models of particle acceleration and energy transport in
  FRIIs. The jets we see at 90 GHz are predominantly in objects with
  atypically bright cm-wavelength jets, though, and it is not clear
  that they tell us what will be seen when more typical FRII sources
  are observed with higher sensitivity.

\item Hotspots have been known for some time to be bright sources at
  90 GHz and beyond, and are clearly detected in all of our targets
  that show compact, bright hotspots at lower frequencies. This is
  unsurprising given the known detections of hotspots (including
  several in our sample) at infrared, optical and X-ray wavelengths in
  synchrotron emission. The wide variety of spectral indices seen
  between 8 and 90 GHz indicates that hotspots have significantly
  different electron energy spectra by energies corresponding to
  90-GHz synchrotron emission.

\item Emission from the lobes is not universal but is common in our
  sample of high-surface-brightness targets; it comes preferentially
  from high-surface-brightness regions of lower-frequency emission
  close to the hotspots. Spectral indices in the lobes between cm and
  mm wavelengths are typically steep, $\alpha > 1.0$, and so
  qualitatively the properties of the lobes are consistent with the
  standard spectral ageing model. In a few cases we have been able to
  show quantitative consistency with a Jaffe \& Perola (1973: J-P)
  electron spectrum, though in at least one case (3C\,123) we have
  argued elsewhere that reacceleration in the lobes may be required to
  produce the high-frequency emission.
\end{itemize}

\subsection{Prospects for ALMA and the EVLA}

The observations summarized in this paper give a preview of the
components that ALMA may be expected to detect when it observes FRII
radio galaxies at frequencies around 90 GHz. However, ALMA will have
instantaneous sensitivity an order of magnitude higher than BIMA,
substantially better $uv$ plane coverage, and higher angular
resolution. The increase in our capability to image mm-wave structures
in extragalactic radio sources will be comparable to that obtained at
cm wavelengths when the VLA first became available at the end of the
1970s. Our work here has shown that there are questions that only ALMA
can answer regarding the detailed spectral structure of radio
galaxies. Since the synchrotron spectra of lobes, particularly near
the hotspots, appear to be cutting off in the mm regime, ALMA will
provide the best possible leverage on models of spectral ageing and
therefore of energy transport and particle dynamics in the lobes. ALMA
should also be able to probe the high-frequency structure of FRII jets
for the first time. We emphasise however that these projects will
require us to combine mm observations with observations at lower
frequencies with instruments like the (expanded) VLA, bearing in mind
that the VLA will have greatly increased sensitivity at frequencies of
tens of GHz once the EVLA upgrade is complete. Higher ALMA frequencies
will be of value in studies of hotspots but will probably not probe
deep into the lobes, and the increasingly small ALMA field of view at
high frequencies will probably prove restrictive in this respect. The
requirement for complementary observations at lower radio frequencies
is a motivation for studies of equatorial samples of lobe-dominated
radio galaxies in the run-up to the ALMA era (e.g. Best, R\"ottgering
\& Lehnert 1999).

\begin{figure*}
\epsfxsize 12cm
\epsfbox{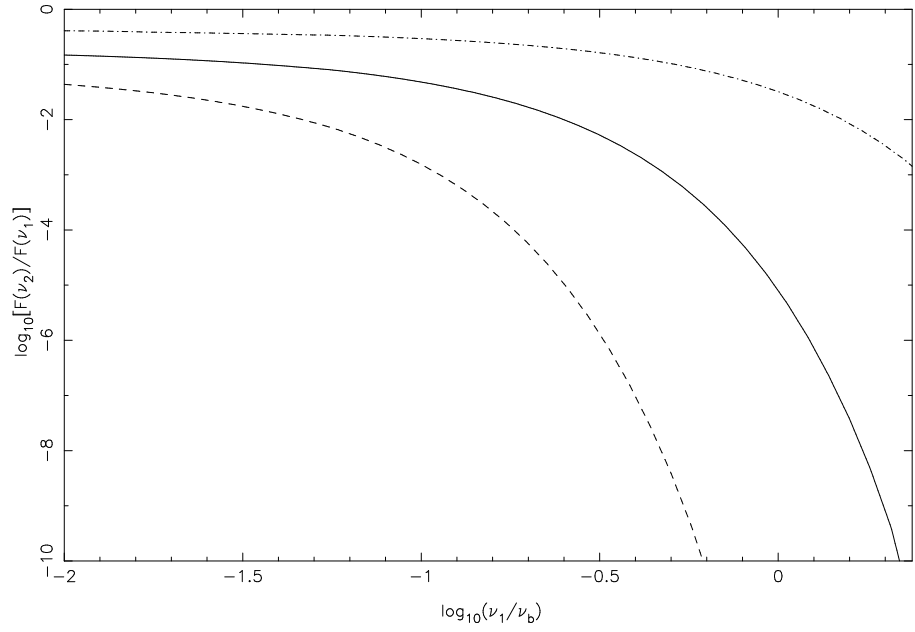}
\caption{The ratio of flux densities at two source-frame frequencies
  $\nu_1$ and $\nu_2$ as a function of $\nu_1/\nu_{\rm b}$, where
  $\nu_{\rm b}$ is the break frequency defined in eq.\ \ref{nub}. A
  Jaffe \& Perola (1973) electron energy spectrum is assumed. The
  dashed line corresponds to $\log_{10}(\nu_2/\nu_1)=1.5$, the solid
  line to $\log_{10}(\nu_2/\nu_1)={1.0}$ and the dotted line to
  $\log_{10}(\nu_2/\nu_1)={0.5}$. The low-frequency spectral index
  (`injection index') $\alpha
  = 0.7$ in this plot.}
\label{jpplot}
\end{figure*}

One intriguing possible use of ALMA is in the direct detection of the
expected Sunzaev-Zel'dovich (S-Z) effect from radio galaxy lobes
(En\ss lin \& Kaiser 2000; Colafrancesco 2008). Since X-ray
inverse-Compton emission from the lobes of radio galaxies is now
routinely detected (e.g. Kataoka \& Stawarz 2005; Croston \etal\ 2005) we
know the electron content of radio lobes well enough to make
predictions at the level of individual radio galaxies, and signatures
of the S-Z effect may well in principle be detectable at mm
wavelengths, with the possibility of providing information about the
low-energy electron spectrum (Colafrancesco 2008). Apart from the
weakness of the signal, the major difficulty is the synchrotron
emission at these wavelengths. Our work in this paper shows that
emission from cores and hotspots is likely to be unavoidable, at least
in active radio galaxies as opposed to relics. However, some hope is
provided by the (limited) evidence that spectral ageing may be
operating as expected at high frequencies in these lobes. In the (most
favourable) case of effective pitch angle scattering (i.e., a J-P
spectrum), assuming losses dominated by a constant, fully tangled
magnetic field of strength $B$, the maximum Lorentz factor in the
electron spectrum is
\begin{equation}
\gamma_{\rm c} = {3m_{\rm e}c\over {4\sigma_{\rm T}}} {1\over {Ut}}
\end{equation}
(Pacholczyk 1970) where $\sigma_{\rm T}$ is the Thomson cross-section,
$m_{\rm e}$ is the mass of the electron, $U$ is the energy
density in the magnetic field ($U = B^2/2\mu_0$) and $t$ is the ageing
time. This energy corresponds to a characteristic `break' frequency given by
\begin{equation}
\nu_{\rm b} = {3 \over {4\pi}} \gamma_{\rm c}^2 {eB \over m_{\rm e}}
\end{equation}
(Longair 1994) where $e$ is the charge on the electron, and so
\begin{equation}
\nu_{\rm b} = {243\pi \over 4} {{m_{\rm e}^5 c^2}\over {e^7\mu_0^2 B^3 t^2}}
\label{nub}
\end{equation}
(Leahy 1991). As shown graphically by Jaffe \& Perola (1973) and Leahy
(1991), the synchrotron spectrum rapidly steepens at frequencies at
and around $\nu_{\rm b}$. Because the steepening in the J-P spectrum
is monotonic, there is a unique value of the break frequency that
produces a flux density ratio $R$ between two observed frequencies
$\nu_1$, $\nu_2$ (s.t. $\nu_2/\nu_1 = f$) if the spectrum is a power
law at frequencies $\ll \nu_{\rm b}$ with a spectral index (the `injection
index') $\alpha$ and $-\log(R)/\log(f) > \alpha$.
Fig.\ \ref{jpplot} shows a plot of the log of $R$ as a function of
$\nu_1/\nu_{\rm b}$ for various values of $f$ and a plausible choice
of $\alpha$. Required values of the break frequency can be read from
this plot. For example, if we know that the flux density at $\nu_1 =
8.4$ GHz (source frame) is 1 Jy, which is at the high end of what is
observed for a typical 3CRR source, and we require the flux density at
$\nu_2 = 90$ GHz to be $<1$ $\mu$Jy (i.e., well below any detectable
S-Z effect), $\log_{10}(f) \approx 1$, $\log_{10}(R) = -6$ and Fig.\
\ref{jpplot} shows that $\nu_{\rm b} \approx \nu_1$. We can then use
eq.\ \ref{nub} to find the required age of the plasma for a given
magnetic field strength. For typical observed magnetic fields $\sim 1$ nT
(Croston \etal\ 2005) we require lobe ages (i.e. times since particle
acceleration) of $\ga 2\times 10^7$ years, which will be achieved only
in large, old radio galaxies; ages comparable to this are however
found in some spectral ageing studies (e.g. Alexander \& Leahy 1987).
Because $t \propto 1/\sqrt \nu_b$, Fig.\ \ref{jpplot} shows that these
conclusions only change weakly if we are searching for the S-Z effect
at higher frequencies. However, there is a strong dependence on the
magnetic field strength, so more powerful radio galaxies with higher
magnetic field strengths may be better targets. Large radio galaxies
are in general the best targets for S-Z detections, both because
spectral ageing will have had the greatest effect and because the
integrated S-Z decrement (for a given low-energy electron spectrum) is
essentially proportional to the energy stored in the lobes, but in
practice there will be a tradeoff between these factors and the
requirement that the source be well imaged in ALMA's small field of
view. Nevertheless we consider that S-Z detection experiments, while
challenging, are not completely ruled out by what our observations
have shown us about the nature of radio galaxies at ALMA frequencies.

We emphasise that the calculation we have carried out here assumes
that the J-P electron spectrum is a completely accurate description of
the lobes. If there is even a very small amount of local particle
reacceleration in radio galaxy lobes, or (equivalently) rapid
transport of high-energy electrons from the hotspots, then the
flat-spectrum radiation produced will completely dominate the mm-wave
emission from the lobes in `old' regions of plasma far from the
hotspots (potentially in the manner we saw in our observations of
3C\,123; Looney \& Hardcastle 2000). Another way of interpreting the
calculations above is that they show ALMA's ability to provide a very
sensitive probe of the degree to which local particle acceleration is
important in lobes. Thus deep ALMA observations of old radio lobes are
likely to produce interesting results whether they detect the S-Z
effect or not.

\section*{Acknowledgements}

We thank Mel Wright for providing us with the calibrated Cygnus A data
and images, and the referee, Prof. D.J. Saikia, for constructive
comments that have helped us improve the paper. MJH thanks the Royal
Society for a research fellowship. LWL acknowledges support from the
Laboratory for Astronomical Imaging at the University of Illinois and
NSF grant AST 0228953. The National Radio Astronomy Observatory is a
facility of the National Science Foundation operated under cooperative
agreement by Associated Universities, Inc.

\clearpage


\clearpage

\end{document}